\documentstyle[aps,epsf,epsfig,preprint]{revtex}

\newcommand{\bq}{\begin{equation}}
\newcommand{\eq}{\end{equation}}
\newcommand{\bqn}{\begin{eqnarray}}
\newcommand{\eqn}{\end{eqnarray}}
\newcommand{\nb}{\nonumber}
\newcommand{\lb}{\label}

\begin{document}
\title{  Kink Stability of Isothermal Spherical  
Self-Similar Flow  Revisited} 
\author{ Anzhong Wang  \thanks{ E-mail: Anzhong$\_$Wang@baylor.edu}}
\address{ CASPER, Physics Department, Baylor University, 101 Bagby Avenue, Waco,
TX76706}
\author{Yumei Wu \thanks{ E-mail: Yumei$\_$Wu@baylor.edu; yumei@im.ufrj.br}}
\address{ Mathematics Department, Baylor University, Waco,
TX76798\\
and\\
Institute of Mathematics, the Federal University of Rio de Janeiro,  
Caixa Postal 68530, CEP 21945-970, Rio de Janeiro, RJ, Brazil}

\date{\today }

\maketitle


\begin{abstract}

The problem of kink stability of isothermal spherical self-similar flow in 
newtonian gravity is revisited. Using distribution theory we first develop 
a general formula of perturbations, linear or non-linear, which consists of 
three   sets of differential equations, one in each side of the 
sonic line and the other  along it. By solving  the equations along 
the  sonic line we find explicitly the spectrum, $k$, of the perturbations, 
whereby we obtain the stability criterion for the self-similar solutions. 
When the solutions are smoothly across the sonic line, our results reduce 
to those of Ori and Piran. To show such obtained perturbations can be matched 
to the ones in the regions outside the sonic line, we  study  the linear 
perturbations in the external region of the sonic line (the ones in the 
internal region are identically zero), by taking the solutions obtained 
along the line as the boundary conditions. After properly imposing other 
boundary conditions at spatial infinity, we are able to show that linear 
perturbations, satisfying all the boundary conditions, exist and do not 
impose any additional conditions on $k$.  As a result, the complete treatment 
of perturbations in the whole spacetime does not alter the spectrum obtained 
by considering  the perturbations only along the sonic line.

\end{abstract}

\vspace{.7cm}




\section{Introduction}

\renewcommand{\theequation}{1.\arabic{equation}}
\setcounter{equation}{0}

In this paper we re-consider the problem of kink instability
of isothermal spherical  self-similar flow in newtonian gravity. This 
problem was first considered by Ori and Piran \cite{OP88} and found
that a large class of solutions was not stable against kink perturbations.

The kink modes result from the existence of sonic lines (points), 
along which discontinuities of (higher order) derivatives of some  
physical quantities can propagate.  Such weak discontinuities cannot occur
spontaneously. They are always the results of some singularities of the initial
and/or boundary conditions. For example, they may occur due to 
the presence of angles on the surface of a body past which the flow 
takes place. In this case the first spatial derivatives of the velocity
are discontinuous. They may also be formed when the curvature of the 
surface of the body is discontinuous (but without angles), in which 
the second spatial derivatives of the velocity are discontinuous.
In addition, any singularity in the time variation of the flow
also results in a non-steady weak discontinuity.
The instability of such discontinuities is characterized by the
divergence of them, and the blow-up may imply the formation of shock 
waves \cite{CF48}. 

An example that discontinuities of derivatives can propagate along a sonic
line is given by the linear perturbation, $\delta\varphi(\tau, x) =
\varphi_{1}(x)e^{k\tau}$, of a massless scalar field in 
$2+1$ gravity, which satisfies the    equation \cite{HW02},
 \bq
 \lb{lp}
 x\left(1-x\right){\varphi_{1}}'' + \frac{1}{2}\left[(1+2k) -x(3+2k)\right]
 {\varphi_{1}}' - \frac{1}{2} k {\varphi_{1}} = f(x),
 \eq
where a prime denotes the ordinary differentiation with respect to
the indicated argument, $f(x)$ is a smooth function of $x$, and $x
= 1$ is the location of the sonic line \footnote{Note the different notations
used here and those in \cite{HW02}. In particular, here we use $x$ 
in the place of $z$ of \cite{HW02}.}. From the above equation we can see that
it is possible for $\varphi_{1}$ to have discontinuous derivatives
across the sonic line $x =1$. In fact, assume that
$\varphi_{1}$ is continuous across $x = 1$ (but not its
first-order derivative), we can   write it in the form
 \bq
 \lb{lpa}
 \varphi_{1}(x) = \varphi_{1}^{+}(x)H(x-1)
 + \varphi_{1}^{-}(x)\left[1 - H(x-1)\right],
 \eq
where $H(x-1)$ denotes the Heavside (step) function, defined as
 \bq
 \lb{lpb}
 H(x-1) = \cases{1, & $x \ge 1$,\cr
 0, & $x < 1$.\cr}
 \eq
Then, we find that
 \bqn
 \lb{lpc}
 {\varphi_{1}}' &=& {
 \varphi_{1}^{+}}' H(x-1)
 + {\varphi_{1}^{-}}'\left[1 - H(x-1)\right],\nb\\
{\varphi_{1}}'' &=& {\varphi_{1}^{+}}'' H(x-1)
 + {\varphi_{1}^{-}}''\left[1 - H(z-1)\right] +
 \left[{\varphi_{1}}'\right]^{-}\delta(x-1),
 \eqn
where $\delta(x-1) \; [= dH(x-1)/dx]$ denotes the Dirac delta function, and
 \bq
 \lb{lpd}
 \left[{\varphi_{1}}'\right]^{-} \equiv
 \lim_{x \rightarrow 1^{+0}}{\left(\frac{d\varphi_{1}^{+}(x)}{dx}\right)}
  - \lim_{x \rightarrow 1^{-0}}{\left(\frac{d\varphi_{1}^{-}(x)}{dx}\right)}.
  \eq
Substituting Eq.(\ref{lpc}) into Eq.(\ref{lp}) and considering the
facts
 \bqn
 \lb{lpe}
 & & H^{m}(x) = H(x),\;\;\; \left[1 - H(x)\right]^{m} = \left[1 -
 H(x)\right],\nb\\
& &  \left[1 - H(x)\right] H(x) =0,\;\; x \delta(x) = 0,
 \eqn
where $m$ is an integer, one can see that Eq.(\ref{lp}) holds also
on the sonic line $x =1$ even when $\left[{\varphi_{1}}'\right]^{-}
\not= 0$. This is because   $(1-x)\left[{\varphi_{1}}'\right]^{-}
\delta(x-1) = 0$ for any finite $\left[{\varphi_{1}}'\right]^{-}$. 
Thus, the quantity
\bq
\lb{lpea}
\left[{\delta\varphi}_{,x}\right]^{-} = 
\left[{\varphi_{1}}'\right]^{-} e^{k\tau},
\eq
has support only along the sonic line,
where $(\;)_{,x} \equiv \partial (\;)/\partial x$.
The above expression shows clearly how the discontinuity, 
$\left[{\delta\varphi}_{,x}\right]^{-}$,
of the first derivative of the perturbation
$\delta\varphi(\tau,x)$ propagates along   the sonic line. When
$Re(k) > 0$ the perturbation grows exponentially as $\tau \rightarrow 
\infty$, and is said unstable with respect to the kink perturbation. 
When $Re(k) = 0$, the perturbation oscillates in time $\tau$, and is usually
also considered as unstable.
When $Re(k) < 0$ the perturbation decays exponentially and is said stable.
 Note that if the discontinuity happened on other places,
say, $x = x_{0} \not= 1$, clearly Eq.(\ref{lp}) would not hold on
$x = x_{0}$, because now $(1-x)\left[{\varphi_{1}}'\right]^{-}
\delta(x-x_{0})\not= 0$. This explains why the discontinuities are
allowed only along sonic lines.

From the above example we can also see that in general one may divide
the whole space into three different regions, $x > x_{c}$,
$x = x_{c}$, and $x < x_{c}$, where $x  = x_{c}$ is the location of the sonic line. 
In each of the three regions, the perturbations, collectively denoted by 
$\delta{y}$, satisfy a set of differential equations,
\bqn
\lb{eqs1}
G^{+}\left(\delta y^{(m)},  \tau, x, k\right) = 0, \;\; 
\left(x \ge x_{c}\right), \\
\lb{eqs2}
G^{-}\left(\delta y^{(m)},  \tau, x, k\right) = 0, \;\; 
\left(x \le x_{c}\right), \\
\lb{eqs3}
G^{(c)}\left(\left[\delta y^{(m)}\right]^{-}, \left[\delta y^{(m)}\right]^{+}, \tau, 
x_{c}, k\right) = 0, \;\; \left(x  = x_{c}\right),
\eqn
where $m = 0, 1, 2, ...,$ and 
\bqn
\lb{eqs4}
\delta y(\tau, x) &=& \delta y^{+}(\tau, x) H\left(x-x_{c}\right)
 + \delta y^{-}(\tau, x)\left[1 - H(x-x_{c})\right],\nb\\
\delta y^{(m)}(\tau, x)  &\equiv&  \frac{\partial^{m} \left(\delta y(\tau, x)\right)}
{\partial x^{m}},\nb\\  
\left[\delta y^{(m)}\right]^{\pm} &\equiv& 
\lim_{x \rightarrow x_{c}^{+}}{\frac{\partial^{m} \left(\delta 
y^{+}(\tau, x)\right)}{{\partial x}^{m}}} 
\pm \lim_{x \rightarrow x_{c}^{-}}{\frac{\partial^{m} \left(\delta y^{-}(\tau, x)\right)}
{{\partial x}^{m}}}.
\eqn
Therefore, in the presence of weak discontinuities the field equations 
actually
have three different parts. Eq.(\ref{eqs3}) describes the 
evolution of the discontinuities, $\left[\delta y^{(m)}\right]^{\pm}$,
along the sonic line, while Eqs.(\ref{eqs1}) and (\ref{eqs2}) describe the 
evolution of $\delta{y}^{\pm}$ in the regions $x > x_{c}$ and $x < x_{c}$.  
Although 
$\delta{y}^{\pm}$ and $\left[\delta y^{(m)}\right]^{\pm}$ satisfy different 
equations, {\em they have to match each other according to Eq.(\ref{eqs4})}.
In addition, to have the perturbations physically acceptable, $\delta {y}^{\pm}$  
usually needs to satisfy some boundary conditions in the regions 
$x > x_{c}$ and $x < x_{c}$, such as, the regularity conditions on 
the symmetry axis, and those at spacelike infinities.

In \cite{OP88} the field equation (\ref{eqs3}) along the sonic line was 
studied, by assuming that $\delta y^{-}(\tau, x) = 0$  in the 
region $|x| < |x_{c}|$. Hanawa and Matsumoto applied such a study to 
a fluid with
a polytrope equation of state, $p = K \rho^{\gamma}$,
where $p$ and $\rho$ denote, respectively, the pressure and mass density of
the fluid, and $K$ and $\gamma$ are two constants \cite{HM00}.
Lately, Harada, and Harada and Maeda generalized such a study to the 
relativistic case  \cite{Har01,HM03}. Since  $\delta y^{-}(\tau, x)$ vanishes 
identically, it is necessary that $\delta y^{+}(\tau, x)$ does not, 
in order to have non-vanishing $\left[\delta y^{(m)}\right]^{-}$.
Then, a natural question is  how the perturbations obtained along 
the sonic line match to the ones in $|x| > |x_{c}|$? 
 
To answer the above question, in this paper
we first solve the equations along the sonic
line for $\delta y_{c}(\tau) \equiv \delta y(\tau, x_{c})$, and find 
explicitly the spectrum, $k$, of the perturbations. 
In \cite{OP88} only the case where the self-similar solutions are smooth 
across the sonic line was studied. Here we generalize such a study to include 
the case where the self-similar solutions can have discontinuous derivatives.
Once $\delta y_{c}(\tau)$ is found, we then solve the equations for the
perturbations $\delta y^{+}(\tau, x)$, by taking $\delta y_{c}(\tau)$ 
and its first derivatives as the  boundary conditions 
[cf. Eqs.(\ref{6.3a}) and (\ref{6.3ab})],
\bqn
\lb{1.12a}
& & \left.\delta y^{+}(\tau, x)\right|_{x = x_{c}} = \delta y_{c}(\tau), \\
\lb{1.12b}
& & \left.\delta y^{+}_{,x}(\tau, x)\right|_{x = x_{c}} = u(\tau).
\eqn
In addition to these, we also impose  conditions  at spatial infinity 
$x = \infty$ [cf. Eq.(\ref{6.3b})],
\bq
\lb{1.12c}
{\mbox{Perturbations grow slower than the background solutions}}.
\eq
In general, conditions (\ref{1.12a}) are already sufficient to determine
all the integration constants. Thus,  conditions (\ref{1.12b}) and (\ref{1.12c})
represent additional restrictions. By carefully
analyzing the perturbations, we are able  
to show that {\em there exist solutions to the perturbation equations 
that satisfy all the boundary conditions (\ref{1.12a}) - (\ref{1.12c}), 
and do not alter the spectrum $k$, obtained by  considering the perturbations
only along the sonic line}. 
 
Specifically, the paper is organized as follows: in Sec. II we review 
some main properties of the self-similar solutions, and pay particular
attention on their asymptotic behavior near the sonic line and at the spatial
infinity. In particular, we work out  the equations that the 
discontinuities of the first-order derivatives of 
the self-similar solutions have to satisfy across the sonic 
line. In Sec. III, by using distribution theory we first develop a general 
formula of perturbations, which are valid for both linear and non-linear 
perturbations. It consists of three different sets, one  in each side
of the sonic line, and the other along it.  Then, in Sec. IV we restrict 
ourselves to the perturbations along the sonic line and find  explicitly
the spectrum, $k$, of the perturbations. 
When the solutions are analytical across
the line, we re-discover the results of Ori and Piran. When they are not,
we find explicitly the stability criterion for those solutions in terms of
$x_{c}$. 
In Sec. V, we consider the linear perturbations $\delta y^{+}(\tau, x)$
in the region  $|x| > |x_{c}|$, by taking the solutions $\delta y_{c}(\tau)$ 
and its derivatives obtained along the sonic line in Sec. IV
 as the boundary conditions. The paper is 
ended  with Sec. VI, where our main conclusions are presented.

\section{Field Equations  and Self-Similar Solutions}
\renewcommand{\theequation}{2.\arabic{equation}}
\setcounter{equation}{0}

The spherical flow of self-gravitating isothermal perfect fluid is 
described by the hydrodynamic equations,
\bqn
\lb{1.1a}
& & v_{,t} + v v_{,r} + \frac{c_{s}^{2}\rho_{,r}}{\rho} 
+ \frac{Gm}{r^{2}} = 0, \\
\lb{1.1b}
& & r^{2} \rho_{,t} + \left(r^{2} v\rho\right)_{,r} = 0,\\
\lb{1.1ca}
& & m_{,t} = - 4\pi r^{2}\rho v,\\
\lb{1.1c}
& & m_{,r} = 4\pi r^{2}\rho,
\eqn
where $v_{,t} \equiv \partial v/\partial t$, 
and $r$, $t$, $c_{s}$, $G$, $v(t,r)$, $\rho(t,r)$ and $m(t,r)$ 
denote, respectively, the radial coordinate,   
the time, the speed of sound, the gravitational constant,  
the velocity, the mass density, and the total mass enclosed within 
the radius $r$ of the perfect fluid at the moment $t$, with
the equation of state  
\bq
\lb{1.2}
p = c^{2}_{s} \rho,
\eq
where $p$ denotes the pressure of the fluid. The boundary conditions
at $r = 0$ are
\bq
\lb{1.3}
v(t,0) = 0, \;\;\; \rho(t,0)   = 0.
\eq

Note that Eqs.(\ref{1.1a})
-  (\ref{1.1c}) are unchanged  under the coordinate transformations,
\bq
\lb{transf}
t = t' + \epsilon, \;\;\;
r = r',
\eq
where $\epsilon$ is a constant. We call such a gauge  as the
{\em Galilean gauge}, and shall be back to it when we consider the linear 
perturbations in Sec. V.

Following Ori and Piran, we first introduce the   dimensionless quantities,
\bqn
\lb{1.4}
& & x = \frac{r}{c_{s} t}, \;\;\;
\tau = - \ln\left(\frac{-t}{t_{0}}\right), \nb\\
& & V = \frac{v}{c_{s}}, \;\;\;
D = \frac{4\pi G\rho}{c_{s}^{2}}r^{2},\;\;\;
M = \frac{Gm}{c_{s}^{2} r},
\eqn
where $t_{0}$ is a dimensional constant with the dimension of length, which,
without loss of generality, shall be set to one. Then,
it can be shown that Eqs.(\ref{1.1a})-(\ref{1.1c}) take the form
\bqn
\lb{1.5a}
& & \left(V - x\right) V_{,x} + E_{,x}  - V_{,\tau} + \frac{1}{x}(M - 2) = 0, \\
\lb{1.5b}
& & 
\left(V - x\right) E_{,x} + V_{,x}  - E_{,\tau}   = 0, \\
\lb{1.5ca}
& & xM_{,x} + M - e^{E} = 0,\\
\lb{1.5c}
& & xM_{,x} + M_{,\tau} - \frac{1}{x}V e^{E}   = 0, 
\eqn
where $E \equiv \ln(D)$. 
Since in this paper we are mainly concerned with the gravitational collapse
of the fluid, from now on we shall restrict ourselves only to the region
where $t \le 0$, in which a singularity is usually developed at the moment
$t = 0$. From Eq.(\ref{1.4}) we can see that this also means $x \le 0$.

Self-similar solutions are given by
\bq
\lb{3.1}
F(\tau, x) = F_{ss}(x),
\eq
where $F \equiv \left\{E, \; M,\; V\right\}$. Setting all the terms
that are the derivatives of $\tau$ be zero in Eqs.(\ref{1.5a})-(\ref{1.5c}),
we find that 
\bqn
\lb{3.2a}
& & \left(V_{ss} - x\right) {V_{ss}}'   + {E_{ss}}'    
+ \frac{1}{x}\left(M_{ss} - 2\right) = 0, \\
\lb{3.2b}& & 
\left(V_{ss} - x\right) {E_{ss}}' + {V_{ss}}'      = 0, \\
\lb{3.2c}
& & x{M_{ss}}' + M_{ss} - e^{E_{ss}} = 0,\\
\lb{3.2ca}
& & x{M_{ss}}'   - \frac{1}{x} e^{E_{ss}} V_{ss} = 0.
\eqn
The boundary conditions of Eq.(\ref{1.3}) become
\bq
\lb{1.3a}
V_{ss}(0) = 0 = D_{ss}(0).   
\eq
From Eqs.(\ref{3.2c}) and (\ref{3.2ca}) we find that
\bq
\lb{mass}
M_{ss} = e^{E_{ss}}\left(1 - \frac{V_{ss}}{x}\right).
\eq

The solutions of these equations have been studied extensively 
\cite{ss,shu77,ws85,hun86}. In particular, it was shown that they must pass a 
sonic line at $x = x_{c}$, defined by \footnote{The line 
$V_{ss}(x_{c}) - x_{c} = -1$ is also a sonic line, with the segment
$t \le 0$ being saddle. However, this part is physically irrelevant, so
in this paper we shall not consider it.}
\bq
\lb{3.3}
V_{ss}(x_{c}) - x_{c} = 1,
\eq
which divides the (t,r)-plane into two regions: the interior, 
$|x| < |x_{c}|$, where $V_{ss} - x - 1 < 0$, and the exterior, $|x| > |x_{c}|$,
where $V_{ss} - x - 1 > 0$, as shown in Fig. 1. The motion of inwardingly
moving perturbations is given by $dr/dt = v(t,r) - c_{s}$, or
\bq
\lb{3.4}
\frac{dx}{dt} = - \frac{1}{|t|}\left(V_{ss} - x - 1\right) = 
\cases{ < 0, & $|x| > |x_{c}|$,\cr
= 0, & $|x| = |x_{c}|$, \cr
> 0, & $|x| < |x_{c}|$, \cr}
\eq
where $t \le 0$. Thus, in the exterior all the perturbations, 
including those
that move inward with respect to the fluid, are dragged outward in terms of
$x$, while in the interior perturbations can move in both directions. As
a results, perturbations cannot penetrate from the exterior into interior
\cite{OP88}. this observation is important when we consider the kink 
perturbations.

 \begin{figure}[htbp]
 \begin{center}
 \label{fig1}
 \leavevmode
  \epsfig{file=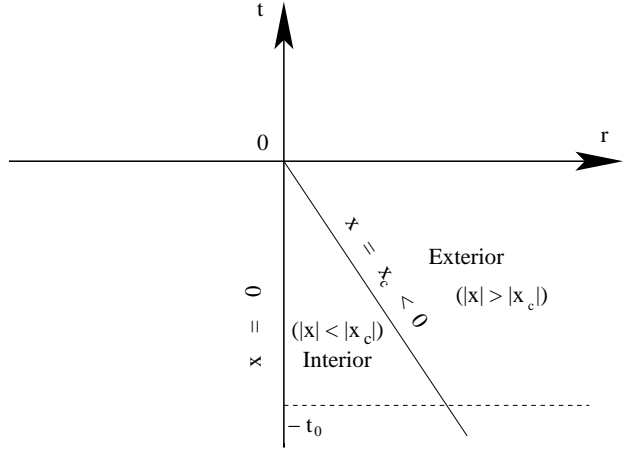,width=0.5\textwidth,angle=0}
 \caption{ The ($t, r$)-plane. The line $x = x_{c} < 0$ represents the   
 sonic line. The whole plane is divided into three regions:  
 $\Omega^{+} \equiv \left\{t, r: |x| \ge |x_{c}|\right\}$;
 $\Sigma \equiv \left\{t, r: x = x_{c}\right\}$;
 and $\Omega^{-} \equiv \left\{t, r: |x| \le |x_{c}|\right\}$.} 
 \end{center}
 \end{figure}

Assuming that $F_{ss}(x)$ is at least $C^{2}$ in $\Omega^{\pm}$,
it can be shown that the  solutions near the sonic line $ x = x_{c}$ 
are given by \cite{shu77,hun86}
\bqn
\lb{3.5}
V_{ss}\left(x\right) &=& \left(1 + x_{c}\right) 
+ \frac{ 1 +  x_{c}}{x_{c}}\left(x - x_{c}\right)
- \frac{1 + x_{c}}{2 x^{2}_{c}} \left(x - x_{c}\right)^{2} +
R_{V}\left(x, x_{c}\right),\nb\\
E_{ss}\left(x\right) &=& \ln\left(-2x_{c}\right)
- \frac{1 +x_{c}}{x_{c}}\left(x - x_{c}\right) 
+ \frac{1 + x_{c}}{x^{2}_{c}} \left(x - x_{c}\right)^{2} 
+ R_{E}\left(x, x_{c}\right),\nb\\
M_{ss}\left(x\right) &=& 2 - \frac{2\left(1  + x_{c}\right)}
{x_{c}}\left(x - x_{c}\right)
+\frac{\left(1 + x_{c}\right)\left(2 + x_{c}\right)}{x^{2}_{c}} 
\left(x - x_{c}\right)^{2} 
+ R_{M}\left(x, x_{c}\right),
\eqn
for type 1, and
\bqn
\lb{3.6}
V_{ss}\left(x\right) &=& \left(1 + x_{c}\right) 
- \frac{ 1}{x_{c}}\left(x - x_{c}\right)
- \frac{x^{2}_{c} + 5x_{c} + 5}
{2 x^{2}_{c}\left(2x_{c} + 3\right)} \left(x - x_{c}\right)^{2} 
+ R_{E}\left(x, x_{c}\right),\nb\\
E_{ss}\left(x\right) &=& \ln\left(-2x_{c}\right)
+ \frac{1 }{x_{c}}\left(x - x_{c}\right)
+ \frac{ x^{2}_{c} - 2}{2 x^{2}_{c}\left(2x_{c} + 3\right)}
 \left(x - x_{c}\right)^{2} + R_{E}\left(x, x_{c}\right),\nb\\
M_{ss}\left(x\right) &=& 2 - \frac{2\left(1  + x_{c}\right)}{x_{c}}
\left(x - x_{c}\right)
+\frac{2 + x_{c}}{x^{2}_{c}} \left(x - x_{c}\right)^{2} 
+ R_{M}\left(x, x_{c}\right),
\eqn
for type 2, where $R_{F}\left(x, x_{c}\right)$ denotes the errors of the 
corresponding expansion. Note that when $x_{c} = -2$ the two types coincide. 
For our present purpose, we can consider this degenerate case as a 
particular one of any of the two types.

The qualitative behavior of the solutions near the critical point depends
on the values of $x_{c}$: (a) When $|x_{c}| < 1$, the critical point is a 
{\em saddle}, and only two solutions pass through it, one for each type. 
(b) When $|x_{c}| > 1$, the critical point is a {\em node} \cite{ws85}. 
In this case,
the directions that the solutions pass the critical point are further
classified into {\em primary} and {\em secondary} directions. In particular,
one of the directions, type 1 solutions for $1 <|x_{c}| < 2$ and type 
2 solutions for $|x_{c}| > 2$, is called the secondary direction, while the
other the primary direction. Only one solution can pass a node in the 
secondary direction, but an infinite number of solutions pass it along the 
primary direction \cite{hun86}. Physically relevant solutions are these
across the sonic line through nodes [cf. Fig. 2]. Thus, in the following we shall 
consider only the case where $x_{c} \le -1$.

 \begin{figure}[htbp]
 \begin{center}
 \label{fig2}
 \leavevmode
  \epsfig{file=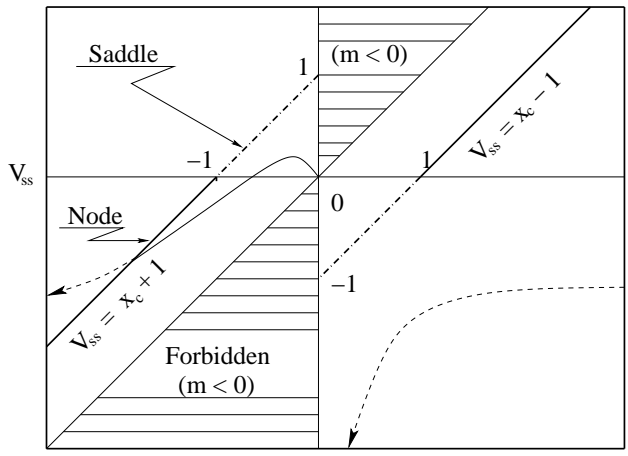,width=0.5\textwidth,angle=0}
 \caption{ The ($V_{ss}, x$)-plane. The region bound by the lines
 $V_{ss}(x) = x $ and $x = 0$ is excluded, because in this region
 the mass of the fluid is negative, $m < 0$. The lines with arrows represent 
 a possible trajectory of the fluid.  } 
 \end{center}
 \end{figure}

It is possible to match solutions with different types across the critical
point. But in this case the matching is only $C^{0}$, that is, $F_{ss}(x)$ 
is only continuous, and  its first-order derivatives normal to the surface $x  = x_{c}$
are usually not \cite{ws85}. Ori and Piran considered only the 
case where the matching is smooth, that is, solutions that belong to the same
type in both sides of the sonic line. In this paper, we shall consider all 
the possibilities.

Dividing the whole plane into three regions, $\Omega^{\pm}$ and $\Sigma$,
where $\Omega^{+} \equiv \left\{t, r: |x| \ge |x_{c}|\right\}$,
$\Sigma \equiv \left\{t, r: x = x_{c}\right\}$, and
$\Omega^{-} \equiv \left\{t, r: |x| \le |x_{c}|\right\}$, we can write
$F_{ss}(x)$ as,
\bq
\lb{3.7}
F_{ss}(x) = \cases{F^{+}_{ss}(x), & $ x \in \Omega^{+}$,\cr
{{F}_{c}}, & $ x \in \Sigma$,\cr
F^{-}_{ss}(x), & $ x \in \Omega^{-}$.\cr}
\eq
Since $F_{ss}$ is $C^{0}$ across the surface $x = x_{c}$, we must have
\bqn
\lb{3.8}
E^{+}_{ss}\left(x_{c}\right) &=& E^{-}_{ss}\left(x_{c}\right)
= \ln\left(-2x_{c}\right),\nb\\
M^{+}_{ss}\left(x_{c}\right) &=& M^{-}_{ss}\left(x_{c}\right)
= 2,\nb\\
V^{+}_{ss}\left(x_{c}\right) &=& V^{-}_{ss}\left(x_{c}\right)
= 1 + x_{c},
\eqn
as we can see from Eqs.(\ref{3.5}) and (\ref{3.6}), this is indeed the case.
On the other hand, from Eqs. (\ref{mass}), 
(\ref{3.5}) and (\ref{3.6}) we find that ${M_{ss}}'$
is also continuous and given by
\bq
\lb{3.9}
{M^{+}_{ss}}'\left(x_{c}\right) = {M^{-}_{ss}}'\left(x_{c}\right)
= - \frac{2(1 + x_{c})}{x_{c}}.
\eq
Taking the limits  $x \rightarrow x_{c}^{ \pm 0}$ in  
Eqs.(\ref{3.2a}) - Eqs.(\ref{3.2ca}) and then subtracting them, we find
that
\bq
\lb{3.10}
\left[{V_{ss}}'\right]^{-} + \left[{E_{ss}}'\right]^{-}= 0,\;\;
(x = x_{c}).
\eq
 Note that in writing Eq.(\ref{3.10}) we had used Eqs.(\ref{3.3})
 and (\ref{3.8}).

As assumed previously, $ F_{ss}(x)$ is  at least  $C^{2}$ in $\Omega^{\pm}$,
we can take  derivative of Eqs.(\ref{3.2a}) - (\ref{3.2ca}) with 
respect to $x$, and obtain   
\bqn
\lb{3.12a}
& & \left(V_{ss}  - x\right) {V_{ss} }'' + {E_{ss} }''  
+ \left({V_{ss} }' - 1\right) {V_{ss} }'     
+ \frac{1}{x}{M_{ss} }' 
- \frac{1}{x^{2}}\left(M_{ss}  - 2\right) = 0, \\
\lb{3.12b}& & 
\left(V_{ss}  - x\right) {E_{ss} }'' + {V_{ss} }'' 
+ \left({V_{ss} }' - 1\right) {E_{ss} }'    = 0, \\
\lb{3.12c}
& & x{M_{ss} }'' + 2{M_{ss} }' - e^{E_{ss} }{E_{ss} }' 
= 0,\\
\lb{3.12ca}
& & x{M_{ss} }'' + {M_{ss} }' 
- \frac{1}{x}e^{E _{ss}}\left({V _{ss}}' 
+ V_{ss} {E _{ss}}'\right)
+ \frac{1}{x^{2}}e^{E _{ss}} V _{ss} = 0,
\eqn
where the quantities $F_{ss}(x)$ in the above equations 
should be understood as  $F^{+}_{ss}(x)$ in $\Omega^{+}$ and
$F^{-}_{ss}(x)$ in $\Omega^{-}$. 
The subtraction of the limits  $x \rightarrow x_{c}^{\pm 0}$ of   
Eqs.(\ref{3.12a}) - Eqs.(\ref{3.12ca}) yields
\bqn
\lb{3.13a}
& & \left[{V_{ss}}''\right]^{-} + \left[{E_{ss}}''\right]^{-}
- \left[{V_{ss}}'\right]^{-}\left(1  - \left[{V_{ss}}'\right]^{+}\right)
=0,\\
\lb{3.13b}
& & \left[{V_{ss}}''\right]^{-} + \left[{E_{ss}}''\right]^{-}
- \left[{E_{ss}}'\right]^{-} +
\frac{1}{2}\left(\left[{E^{}_{ss}}'\right]^{+}\left[{V^{}_{ss}}'\right]^{-}
+ \left[{E^{}_{ss}}'\right]^{-}\left[{V^{}_{ss}}'\right]^{+}\right)
 = 0,\\
\lb{3.13c}
& &  \left[{M_{ss}}''\right]^{-} 
+2  \left[{E_{ss}}'\right]^{-} =0,  \;\; (x = x_{c}),
\eqn
where
\bq
\lb{3.11}
\left[{F_{ss}}'\right]^{+} \equiv 
\lim_{x \rightarrow x_{c}^{+0}}{
\left(\frac{d F^{+}_{ss}\left(x\right)}{d
x}\right)} +
\lim_{x \rightarrow x_{c}^{-0}}{
\left(\frac{d F^{-}_{ss}\left(x\right)}{d
x}\right)}.
\eq
Note that in writing the above equations we had used 
Eqs.(\ref{3.8}) and (\ref{3.10}), so that Eqs.(\ref{3.12c})
and (\ref{3.12ca}) yield the same equation, (\ref{3.13c}).
From Eqs.(\ref{3.13a}) and (\ref{3.13b}), on the other hand,
 we find that
\bq
\lb{3.14}
\left[{V^{}_{ss}}'\right]^{-}
\left(\left[{E^{}_{ss}}'\right]^{+} 
- 3 \left[{V^{}_{ss}}'\right]^{+} + 4\right) = 0,
\;\; (x = x_{c}),
\eq
which has the solutions,
\bqn
\lb{3.15a}
& & (a) \;\; \left[{V^{}_{ss}}'\right]^{-} = 0,\\
\lb{3.15b}
& & (b) \;\; \left[{E^{}_{ss}}'\right]^{+} 
- 3 \left[{V^{}_{ss}}'\right]^{+} + 4 = 0.
\eqn
Clearly, Case (a)   requires that the solutions inside and outside
the surface $x = x_{c}$  belong to the same type, and in this case  
from Eqs.(\ref{3.5}), (\ref{3.6}) and (\ref{3.10}) we find that
\bq
\lb{3.15d}
\left[{E^{}_{ss}}'\right]^{+} - 3 \left[{V^{}_{ss}}'\right]^{+} + 4 =
\frac{4\left(2 - |x_{c}|\right)}{|x_{c}|}
\times \cases{1, & Type 1,\cr
-1, & Type 2, \cr} \;\;\; 
\left(\left[{V^{}_{ss}}'\right]^{-} = 0\right).
\eq

\section{  General Formula of Perturbations }
\renewcommand{\theequation}{3.\arabic{equation}}
\setcounter{equation}{0}

In this section
using distribution theory, we shall develop formulas that are valid 
for any kind of perturbations, linear or non-linear, and write
down the equations of perturbations in $\Omega^{\pm}$ and $\Sigma$
separately.

For any given function $F(x, \tau)$
that is $C^{2}$ in $\Omega^{\pm}$ and $C^{0}$ across the surface 
$x = x_{c}$ can be written as\footnote{Note that in $\Omega^{+}\;
(\Omega^{-})$ we now have $x < x_{c}\; (x > x_{c})$.},
\bq
\lb{4.3}
F(x, \tau) = F^{+}(x, \tau)\left[1 - H\left(x - x_{c}\right)\right]
+ F^{-}(x, \tau) H\left(x - x_{c}\right),
\eq
where $F^{+}$ ($F^{-}$) is defined in $\Omega^{+}$ ($\Omega^{-}$), and
\bq
\lb{4.4}
\lim_{x \rightarrow x^{+0}_{s}}{F^{+}(x, \tau)}
= \lim_{x \rightarrow x^{-0}_{s}}{F^{-}(x, \tau)}
\equiv {{F}_{c}}(\tau).
\eq
In the present case, it is clear   that
\bq
\lb{4.4a}
F^{\pm}(x, \tau) = F^{\pm}_{ss}(x) + \delta F^{\pm}(x, \tau),
\eq
where $F^{\pm}_{ss}(x)$ denotes the self-similar solutions
of Eqs.(\ref{3.2a}) - (\ref{3.2ca}) with the boundary conditions
of Eqs. (\ref{1.3a}), (\ref{3.8}) and (\ref{3.9}). 
The perturbations $\delta F^{\pm}(x, \tau)$
satisfy the conditions, 
\bq
\lb{4.4b}
\lim_{x \rightarrow x^{+0}_{s}}{\delta F^{+}(x, \tau)}
= \lim_{x \rightarrow x^{-0}_{s}}{\delta F^{-}(x, \tau)}
\equiv {\delta{F}_{c}}(\tau).
\eq
Then, using Eqs. (\ref{lpe}), (\ref{4.3}), (\ref{4.4}), and the fact
that
\bq
\lb{4.2}
 H\left(x - x_{c}\right)\delta\left(x - x_{c}\right)
= \frac{1}{2} \delta\left(x - x_{c}\right),\;\;\;
\left[1 - H\left(x - x_{c}\right)\right]\delta\left(x - x_{c}\right)
= \frac{1}{2} \delta\left(x - x_{c}\right), 
\eq 
we find
\bqn
\lb{4.5}
& & F_{,x} = {F^{+}}_{,x}\left[1 - H\left(x - x_{c}\right)\right]
+ {F^{-}}_{,x} H\left(x - x_{c}\right),\nb\\ 
& & F_{,xx} = {F^{+}}_{,xx}\left[1 - H\left(x - x_{c}\right)\right]
+ {F^{-}}_{,xx} H\left(x - x_{c}\right)
- \left[F_{,x}\right]^{-} \delta\left(x - x_{c}\right),\nb\\
& & F^{(n)} = {F^{+}}^{(n)}\left[1 - H\left(x - x_{c}\right)\right]
+ {F^{-}}^{(n)} H\left(x - x_{c}\right)
- \sum^{n-2}_{i = 0}{\left[F^{(n-i -1)}\right]^{-} 
\delta^{(i)}\left(x - x_{c}\right)},
\eqn
where
\bqn
\lb{4.6}
& & \left[F^{(n)}\right]^{-} \equiv \lim_{x \rightarrow x_{c}^{+0}}{
\left(\frac{\partial^{n} F^{+}\left(x, \tau\right)}{\partial
x^{n}}\right)} -
\lim_{x \rightarrow x_{c}^{-0}}{
\left(\frac{\partial^{n} F^{-}\left(x, \tau\right)}{\partial
x^{n}}\right)},\nb\\
& & \delta^{(n)}\left(x - x_{c}\right) \equiv
\frac{d^{n}  \delta\left(x - x_{c}\right)}{d x^{n}},\nb\\
& & \int^{\infty}_{-\infty}{F(x, \tau)\delta^{(k)}\left(x - x_{c}\right)
dx} = (-1)^{k}F^{(k)}\left(x_{c}, \tau\right).
\eqn

Inserting Eqs.(\ref{4.3})-(\ref{4.5}) into Eqs.(\ref{1.5a})  -
(\ref{1.5c}), and considering Eq.(\ref{4.2}), we find that
\bqn
\lb{4.7a}
& & \left(V_{ss} - x\right) {\delta V}_{,x}
+ {V_{ss}}_{,x} \delta V + \delta V
{\delta V}_{,x} - {\delta V}_{,\tau}
+ {\delta E}_{,x} + \frac{1}{x} \delta M = 0,\\
\lb{4.7b}
& & \left(V_{ss} - x\right) {\delta E}_{,x}
+ {E_{ss}}_{,x} \delta V + \delta V
{\delta E}_{,x} - {\delta E}_{,\tau}
+ {\delta V}_{,x}   = 0,\\
\lb{4.7ca}
& & x  {\delta M}_{,x}
+   \delta M -  e^{E_{ss}}\left(e^{\delta E}
- 1 \right) = 0,  \\
\lb{4.7c}
& & x  {\delta M}_{,x}
+   {\delta M}_{,\tau} + \frac{1}{x}  e^{E_{ss}}
\left\{V_{ss}\left(1 - e^{\delta E}\right)
- e^{\delta E} \delta V \right\} = 0.  
\eqn 
 Again,  the quantities $F,\; \delta F$ in the above equations 
should be understood as  $F^{+},\; \delta F^{+}\; \left(F^{-}, 
\; \delta F^{-}\right)$ in $\Omega^{+}\; (\Omega^{-})$.  
The boundary conditions (\ref{1.3}), together with those of Eq.(\ref{1.3a})
require
\bq
\lb{3.12d}
{\delta V^{-}}(t, 0) = 0  = {\delta D^{-}}(t, 0).
\eq

Subtracting the limits $ x \rightarrow x_{c}^{\pm 0}$
of Eqs.(\ref{4.7a}) - (\ref{4.7c}), on the other hand, we find,
\bqn
\lb{4.8a}
& & \left(1 + \delta V_{c}\right) W_{c}(\tau)  
+ U_{c}(\tau)
+ \left[{V_{ss}}'\right]^{-} {\delta V_{c}} = 0, \\
\lb{4.8b}
& & \left(1 + \delta V_{c}\right) U_{c}(\tau)
+ W_{c}(\tau)
+ \left[{E_{ss}}'\right]^{-} {\delta V_{c}} = 0, \\
\lb{4.8c}
& &  \left[{\delta M}_{,x}\right]^{-}    = 0, \;\; (x = x_{c}),
\eqn
where 
\bq
\lb{4.9}
U_{c}(\tau) \equiv \left[{\delta E}_{,x}\right]^{-},\;\;\;
W_{c}(\tau) \equiv \left[{\delta V}_{,x}\right]^{-}, 
\eq
and $\left[{\delta F}_{,x}\right]^{-}$  is defined as
\bq
\lb{4.9a}
\left[{\delta F}_{,x}\right]^{\pm} \equiv 
\lim_{x \rightarrow x_{c}^{+0}}{
\left(\frac{\partial\left(\delta F^{+}\left(x, \tau\right)\right)}{\partial
x}\right)} \pm
\lim_{x \rightarrow x_{c}^{-0}}{
\left(\frac{\partial\left(\delta F^{-}\left(x, \tau\right)\right)}{\partial
x}\right)}.
\eq 
Note that in writing Eqs.(\ref{4.8a}) -(\ref{4.8c}), we used the fact that
\bq
\lb{4.10}
\left. \frac{\partial}{\partial \tau}\delta F^{+}\left(x, \tau\right)
\right|_{x = x_{c}^{+0}}
= \left. \frac{\partial}{\partial \tau}\delta F^{-}\left(x, \tau\right)
\right|_{x = x_{c}^{-0}}.
\eq
In addition, Eqs.(\ref{4.7ca}) and (\ref{4.7c}) give the same equation,
(\ref{4.8c}). Combining Eq.(\ref{3.10}) with Eqs.(\ref{4.8a})-(\ref{4.8c}) 
we find  the following solutions,
\bqn
\lb{4.11a}
& & (i) \;\; {\delta V}_{c} = 0, \;\;\;
U_{c}(\tau) = - W_{c}(\tau), \;\;\;
\left[{\delta M}_{,x}\right]^{-} = 0, \\
\lb{4.11b}
& & (ii) \;\; {\delta V}_{c} = -2, \;\;\;
U_{c}(\tau) = W_{c}(\tau) + 2 \left[{V_{ss}}'\right]^{-}, \;\;\;
\left[{\delta M}_{,x}\right]^{-} = 0, \\
\lb{4.11c}
& & (iii) \;\;  U_{c}(\tau) = - W_{c}(\tau) = \left[{V_{ss}}'\right]^{-}, \;\;\;
\left[{\delta M}_{,x}\right]^{-} = 0, \;\; (x = x_{c}).
\eqn

On the other hand, taking the derivatives of Eqs.(\ref{1.5a})
- (\ref{1.5c}) with respect to $x$, we find that
\bqn
\lb{4.12a}
& & 
\left(V - x\right) V_{,xx} + E_{,xx} - V_{,x\tau}
+ \left(V_{,x} - 1\right) V_{,x} + \frac{1}{x}M_{,x}
- \frac{1}{x^{2}}\left(M - 2\right) = 0,\\
\lb{4.12b}
& & 
\left(V - x\right) E_{,xx} + V_{,xx} - E_{,x\tau}
+ \left(V_{,x} - 1\right) E_{,x}   = 0,\\
\lb{4.12ca}
& & 
 x  M_{,xx} + 2 M_{,x} - e^{E}E_{,x} = 0,\\
 \lb{4.12c}
& & 
 x  M_{,xx} +  M_{,\tau x} + M_{,x} 
 - \frac{1}{x}e^{E}\left(V_{,x} + V E_{,x}\right)
 +  \frac{1}{x^{2}} V e^{E}   = 0.
\eqn
Substituting Eqs.(\ref{4.4a}) - (\ref{4.5}) into the these equations
and using Eq.(\ref{4.2}), we find that each of these equations can be
written in the form,
\bq
\lb{general}
G^{+}(\tau, x)\left[1 - H\left(x - x_{c}\right)\right]
+ G^{-}(\tau, x) H\left(x - x_{c}\right) 
+ G_{c}\left(\tau\right)\delta\left(x - x_{c}\right) = 0,
\eq
which is equivalent to
\bqn
\lb{ga}
G^{+}(\tau, x) = 0, \;\;\; \left(x \in \Omega^{+}\right) \\
\lb{gb}
G^{-}(\tau, x) = 0, \;\;\; \left(x \in \Omega^{-}\right) \\
\lb{gc}
G_{c}\left(\tau\right) = 0, \;\;\; \left(x  \in \Sigma\right).
\eqn
It can be shown that Eqs.(\ref{ga}) and (\ref{gb}) take
the same forms as these given by Eqs.(\ref{4.12a}) - (\ref{4.12c})
after replying $F$ by $F^{\pm}$.
But, on the surface $x = x_{c}$ Eq.(\ref{gc}) gives
\bqn
\lb{4.13}
\left[V_{,x}\right]^{-} + \left[E_{,x}\right]^{-} = 0,\\
\lb{4.13a}
\left[M_{,x}\right]^{-} = 0.
\eqn
Eq.(\ref{4.13a}) is consistent with Eqs.(\ref{3.9}) and
(\ref{4.8c}), while Eq.(\ref{4.13}) together with Eq.(\ref{3.10}) 
yields
\bq
\lb{4.14}
U_{c}(\tau) + W_{c}(\tau) = 0.
\eq
On the other hand,   subtracting the limits $x \rightarrow
x_{c}^{\pm 0}$ of Eqs.(\ref{4.12a})-(\ref{4.12c}), we find
that   
\bqn
\lb{4.15a}
& & \frac{d W_{c}(\tau)}{d \tau}
- \left(1 + {\delta V}_{c}\right) \left[{\delta V}_{,xx}\right]^{-}
- \left[{\delta E}_{,xx}\right]^{-}  
- {\delta V}_{c}\left[{V_{ss}}''\right]^{-} \nb\\
& & \;\;\;\;\;\;\; + \left(1 - \left[{V_{ss}}'\right]^{+} 
- \left[{\delta V}_{,x}\right]^{+}\right)W_{c}(\tau)
- \left[{V_{ss}}'\right]^{-}\left[{\delta V}_{,x}\right]^{+} = 0,\\
\lb{4.15b}
& & \frac{ dU_{c}(\tau) }{d\tau} 
- \left(1 + {\delta V}_{c}\right) \left[{\delta E}_{,xx}\right]^{-}
- \left[{\delta V}_{,xx}\right]^{-}  
- {\delta V}_{c}\left[{E_{ss}}''\right]^{-} \nb\\
& & \;\;\;\;\;\;\; + \frac{1}{2}\left(2 - \left[{V_{ss}}'\right]^{+} 
- \left[{\delta V}_{,x}\right]^{+}\right)U_{c}(\tau)
- \frac{1}{2}\left(\left[{E_{ss}}'\right]^{+}
+ \left[{\delta E}_{,x}\right]^{+}\right)W_{c}(\tau)\nb\\
& & \;\;\;\;\;\;\; - \frac{1}{2}\left(\left[{V_{ss}}'\right]^{-}
\left[{\delta E}_{,x}\right]^{+}
+ \left[{E_{ss}}'\right]^{-} \left[{\delta V}_{,x}\right]^{+}\right)
= 0,\\
\lb{4.15c}
& & x_{c} \left[{\delta M}_{,xx}\right]^{-}
-2x_{c}\left(1 - e^{\delta E_{c}}\right)
 \left[{E_{ss}}'\right]^{-} +2x_{c} e^{\delta E_{c}} U_{c}(\tau) = 0,\\
 \lb{4.15ca}
& &  \left(\left[{E_{ss}}'\right]^{-}+ U_{c}(\tau)\right)\delta V_{c} = 0.
\eqn
Equations (\ref{4.7a}) - (\ref{4.7c}), (\ref{4.11a}) - (\ref{4.11c}),
 and (\ref{4.14}) - (\ref{4.15ca}) 
consist of the full set of differential equations for the
perturbations $\delta F^{\pm}(\tau, x)$ in the whole spacetime, which 
consists of three regions,
$\Omega^{\pm}$ and $\Sigma$.  Note that these equations are exact, and 
so far no approximation has been made. In Sec. V we shall consider their 
linearized perturbations.

\section{Kink Stability}
\renewcommand{\theequation}{4.\arabic{equation}}
\setcounter{equation}{0}

Kink stability is the study of the perturbations of Eqs.(\ref{4.11a}) - 
(\ref{4.11c}), and (\ref{4.14})
- (\ref{4.15ca}) along the critical line $x = x_{c}$.
To solve these   equations for $\delta F^{\pm}(\tau, x)$, in addition 
to (\ref{3.12d}), other boundary conditions need to be given. Following
Ori and Piran, we shall impose the following conditions: Assume that the
perturbations turn on at the moment $t = t_{0}$ or $\tau = 0$ [cf. Fig. 1],
then we require

(A)  the perturbations initially vanish in the interior,
\bq
\lb{4.16a}
\delta F^{-}(\tau = 0, x) = 0,\; x \in \Omega^{-},
\eq

 (B)  the perturbations be continuous everywhere, and in
particular across the sonic line,  
\bq
\lb{4.16b}
\left[{\delta F }\right]^{-} = 0, \;\; \left(x = x_{c}\right),  
\eq

(C)  ${\delta E^{\pm}}_{,x}$ and ${\delta V^{\pm}}_{,x}$  be discontinuous 
at the sonic line, 
\bq 
\lb{4.16c}
U_{c}(\tau) = \left[\delta E_{,x} \right]^{-} \not= 0, \;\;\;
W_{c}(\tau) = \left[\delta V_{,x} \right]^{-} \not= 0, 
\;\; \left(x = x_{c}\right).
\eq
From the above we first note that Eq.(\ref{4.16a}) remains true for all
the $\tau > 0$, because the perturbations cannot penetrate the sonic line 
from exterior into the interior, as we noticed in Sec. II, and 
$\delta F^{-}(x, \tau) = 0$ are indeed solutions of   
Eqs.(\ref{4.7a}) - (\ref{4.7c}) in $\Omega^{-}$. Thus, in the following   
we need to consider only the perturbations in $\Omega^{+}$ as well as 
those  along the sonic line $x = x_{c}$. 

The studies of the perturbations in $\Omega^{+}$
will be considered in the next section, while in the 
rest of this section we shall consider only Eqs.(\ref{4.11a}) - (\ref{4.11c})
and (\ref{4.14}) - (\ref{4.15ca}).
As first noticed by Ori and Piran, the evolutions
of these two sets of equations are separable. In particular,
Conditions (\ref{4.16a}) - (\ref{4.16c}) together with 
Eqs.(\ref{4.11a}) - (\ref{4.11c}) and (\ref{4.14}) - (\ref{4.15ca})
  are already sufficient 
to determine the evolution of $\left[\delta F_{,x} \right]^{-}$ 
uniquely. To show this, let us first notice that 
Conditions (A) and (B) imply 
\bqn
\lb{4.17}
& & \delta F_{c}(\tau) =  \delta F^{-}\left(\tau, x_{c}\right) = 0,\nb\\
& & \left[\delta F_{,x} \right]^{-} = 
\left. \frac{\partial \delta F^{+}\left(\tau, x\right)}
{\partial x}\right|_{x =
x_{c}^{+0}} = \left[\delta F_{,x} \right]^{+}.
\eqn
Then, we can see that only the case of Eq.(\ref{4.11a}) is consistent with
these expressions. Thus, in the following we shall discard the cases 
described by Eqs.(\ref{4.11b}) and (\ref{4.11c}). Hence, one can see 
that Eq.(\ref{4.15ca}) is satisfied identically,
while  from Eqs.(\ref{4.15a}) - (\ref{4.15c}) we find
\bq
\lb{4.18a}
 \frac{dW_{c}}{d\tau} + \frac{1}{4} \left(4 + \left[{E^{}_{ss}}'\right]^{+} 
- 3 \left[{V^{}_{ss}}'\right]^{+} - 4 \left[{V^{}_{ss}}'\right]^{-}\right)
W_{c} - W_{c}^{2} = 0,  
\eq
where in writing this equation we had used Eq.(\ref{4.14}).  To  study
it, let us consider the
two cases defined by Eqs.(\ref{3.15a}) - (\ref{3.15b}) separately.  

{\bf Case A) $\; \left[{V_{ss}}'\right]^{-} = 0$:} In this case the 
internal and external solutions belong to the same type, and we
have  
\bq
\lb{4.19}
{F_{ss}^{+}}'(x_{c}) = {F_{ss}^{-}}'(x_{c}) \equiv {F_{c}}',
\eq
for which Eq.(\ref{4.18a}) becomes
\bq
\lb{4.20}
\frac{dW_{c}}{d\tau} = k W_{c} + W_{c}^{2},\;\;\; 
k \equiv \left(2{V_{c}}' - 1\right).
\eq
This is exactly  the equation  obtained by Ori and Piran \cite{OP88}. 
In particular, it was shown that when ${V_{c}}' \ge 1/2$ the self-similar 
solutions are unstable against the kink perturbations.

{\bf  Case B) $\; \left[{V_{ss}}'\right]^{-} \not= 0$:} This is the case of
Eq.(\ref{3.15b}), from which we find that Eq.(\ref{4.18a}) takes the same
form as that of Eq.(\ref{4.20}), but now with 
\bq
\lb{4.21}
k = \left[{V^{}_{ss}}'\right]^{-}.
\eq
The dynamics of small perturbations are determined by the linear part of
Eq.(\ref{4.20}),
\bq
\lb{4.22}
\frac{dW_{c}}{d\tau} \simeq  k W_{c}, 
\eq 
which has the solution,
\bq
\lb{4.23}
 W_{c}(\tau) \simeq W_{0}e^{k \tau} = \cases{ \pm \infty, & $ k > 0$,\cr
 W_{0}, & $ k = 0$,\cr
 0, & $k < 0$,\cr} \;\; ({\mbox{as}} \; \tau \rightarrow \infty),
\eq 
where $W_{0}$ is an integration constant
and denotes the initial profile of the perturbation.
Therefore, when $k \ge 0$ the corresponding self-similar solutions
are unstable against kink perturbations, and when $k < 0$, they are
stable. To find out the sign of $k$ we need to consider the two cases
where the external solution is Type 1 or 2 separately. 
When it is type 1, then the internal solution must be type 2, and   
then from Eqs.(\ref{3.5}) and (\ref{3.6}) we find
\bq
\lb{4.24}
k= {V_{ss}^{+}}'(x_{c})- {V_{ss}^{-}}'(x_{c})
= \frac{|x_{c}| -2}{|x_{c}|} = \cases{ \ge 0, & $ x_{c} \le -2$,\cr
< 0, & $ -1 > x_{c} > -2$.\cr}
\eq
Thus, for this kind of matching, the self-similar solutions are stable
for $ -1 > x_{c} > -2$, and unstable for $x_{c} \le -2$.
When the external solution is type 2, the internal solution must 
be type 1. Then, the sign of $k$ is just opposite to the last case,
and we conclude that the corresponding self-similar solutions now are unstable
for $ -1 > x_{c} \ge -2$ and stable for $x_{c} < -2$. 

It is interesting to notice that    Eq.(\ref{4.23})  
allows the following general solution \cite{OP88},
\bq
\lb{4.25}
W_{c}(\tau) = \frac{kW_{0} e^{k\tau}}{  {k}  
+ W_{0}\left(1 - e^{k\tau}\right)},
\eq
which can diverge for certain choice of the initial profile $W_{0}$.
Ori and Piran argued that this divergence could be the indication of
the formation of a shock wave.

\section{Linear Perturbations in Region $\Omega^{+}$  }

\renewcommand{\theequation}{5.\arabic{equation}}
\setcounter{equation}{0}

In this section, we shall study the linear perturbations of the
self-similar solutions given in Sec. III in the region $\Omega^{+}$,
where $|x| \ge \left|x_{c}\right|$, with 
$\delta F^{+}\left(\tau, x_{c}\right)$
and $\delta {F^{+}}_{,x}\left(\tau, x_{c}\right)$   obtained in the
last section as the boundary conditions. Let us first write
the perturbations $\delta F^{+}\left(\tau, x \right)$ as
\bq
\lb{6.1}
\delta F^{+}\left(\tau, x\right) =  \epsilon F_{1}(x) e^{k\tau},
\eq
where $\epsilon$ is a very small quantity,  and $F_{1}(x)$  
  denotes perturbations. It is understood that there may have many 
perturbation modes for different   $k$. 
Then, the general linear perturbation will be the
sum of these individual ones.     

To the first order of $\epsilon$,    the substitution 
of Eq.(\ref{6.1}) into Eqs.(\ref{4.7a}) - (\ref{4.7c}) yields
\bqn
\lb{6.2a}
& & \left(V_{ss} - x\right) {V _{1}}'  + { E_{1}}'
+ \left({V_{ss}}' - k\right)  V_{1} + \frac{1}{x}   M_{1} = 0,\\
\lb{6.2b}
& & \left(V_{ss} - x\right) { E _{1}}' +  {V_{1}}'
+ {E_{ss}}'  V_{1} - k E_{1} = 0,\\
\lb{6.2c}
& & x  { M_{1}}' +    M_{1} -  e^{E_{ss}} E_{1} = 0, \\
\lb{6.2d}
& & x  { M_{1}}' +    kM_{1} -  \frac{1}{x}e^{E_{ss}}
\left(V_{ss} E_{1} + V_{1}\right) = 0,  
\;\; \left(x \in \Omega^{+}\right),
\eqn
where,  without causing any confusions, in writing the above equations 
we had dropped the superscript ``+".

Physically acceptable solutions of the above equations must satisfy 
some boundary conditions
at the sonic line $x = x_{c}$ as well as at the spatial infinity $x = \infty$.
At $x = x_{c}$, the boundary conditions are the solutions obtained in the 
last section along $x = x_{c}$.  From Eqs.(\ref{4.8c}), (\ref{4.14}), 
(\ref{4.17}) and (\ref{4.23}) we find that these conditions read
\bqn
\lb{6.3a}
& & V_{1}\left(x_{c}\right) = E_{1}\left(x_{c}\right) = M_{1}\left(x_{c}\right)
= 0, \\
\lb{6.3ab}
& & {V_{1}}'\left(x_{c}\right) = - {E_{1}}'\left(x_{c}\right) =
W_{0}, \;\;\;
 {M_{1}}'\left(x_{c}\right) = 0,\;\; \left(x = x_{c}\right),
\eqn
where $W_{0}$ is the integration constant given in Eq.(\ref{4.23}).
In the neighborhood of $x = x_{c}$, the functions
$F_{ss}(x)$ behave as those given by Eqs.(\ref{3.5}) and (\ref{3.6}).

On the other hand, at the spatial infinity $x = \infty$ 
(or $r = \infty$), we require that the perturbations $F_{1}(x)$ should not 
grow faster than the background solutions $F_{ss}(x)$, that is,
\bq
\lb{6.3b}
F_{1}(x) {\mbox{ grows slower than }} \; F_{ss}(x),
\;\; {\mbox{ as}}\; x \rightarrow \infty,
\eq
where $F_{1} \equiv \left\{E_{1}, V_{1}, M_{1} \right\}$. 
 The asymptotic behavior of $F_{ss}(x)$ is given by \cite{ws85},
 \bqn
 \lb{6.5}
V_{ss}(x) &=& V_{\infty} - \frac{\left(D_{\infty} - 2\right)}{x}
+ \frac{V_{\infty}}{x^{2}} 
+ \frac{\left[4 V_{\infty} + \left(D_{\infty} - 2\right)
\left(D_{\infty} - 6\right)\right]}{6 x^{3}}
+ O\left(x^{-4}\right),\nb\\
E_{ss}(x) &=& \ln\left(D_{\infty}\right) 
-   \frac{\left(D_{\infty} - 2\right)}{2 x^{2}}
+ O\left(x^{-4}\right), 
\eqn
as $x \rightarrow \infty $, where  $D_{\infty}$ and $V_{\infty}$ 
are two arbitrary constants. 

The asymptotic behavior of $M_{ss}(x)$ as $x \rightarrow x_{c}$
and $x \rightarrow \infty$ can be obtained from 
Eqs.(\ref{mass}), (\ref{3.5}), (\ref{3.6})  and (\ref{6.5}).

Since $F_{ss}(x)$ is well-defined in the region $x \in \left(-\infty,
-|x_{c}|\right)$,  one can see that there always exist solutions
to Eqs.(\ref{6.2a})- (\ref{6.2c}), which in general has three free
parameters. However, Eq.(\ref{6.2d}) represents a constraint, and
shall reduce the number of the free parameters from three to two. 
On the other hand, one can show that $M_{1}(x_{c}) = 0$ holds, provided that
$V_{1}(x_{c}) = E_{1}(x_{c}) = 0$ is true. Thus,  
the boundary conditions given by Eq.(\ref{6.3a}) are already sufficient
to determine  the two free parameters uniquely. Then, an important
question  rises: Do  Eqs.(\ref{6.3ab}) 
and (\ref{6.3b}) give further restrictions on these parameters?
In particular, once all the conditions of Eqs.(\ref{6.3a}) - (\ref{6.3b})
are taken into account, do the values of the spectrum $k$ still remain 
the same as those given by Eq.(\ref{4.20}) or (\ref{4.21})? 
If not, is it possible that $Re(k)$ cannot be positive any more, 
and that all the unstable modes found in the last section are due to the 
incomplete treatment of the problem? 

Before addressing these questions, we first consider the problem
concerning gauge modes. Under the coordinate transformations 
(\ref{transf}), we have
\bqn
\lb{gauge}
& & \tau \rightarrow \tau + \epsilon e^{\tau},
\;\;\;
x \rightarrow x\left(1 + \epsilon  e^{\tau}\right),\nb\\
& & F_{ss}(x) \rightarrow F_{ss}(x) + \epsilon F^{(g)}_{1}(x) e^{\tau},
\eqn
where
\bq
\lb{gaugeb}
F^{(g)}_{1}(x) \equiv x {F'}_{ss}(x).
\eq
Comparing it with Eq.(\ref{6.1}) we find that the gauge mode corresponds
to the  case where $k = 1$. This mode is not physical, and completely 
due to the gauge transformations (\ref{transf}). 
Thus, in the following  we shall discard this case and consider 
only the case where $k \not= 1$. Then, from Eqs.(\ref{6.2c}) and 
(\ref{6.2d}) we find 
\bq
\lb{6.6}
M_{1} = - \frac{1}{(1-k)x}\left[\left(V_{ss} - x\right) E_{1} 
+ V_{1}\right] e^{E_{ss}}.
\eq
Combining it with Eqs.(\ref{6.2a}) and (\ref{6.2b}) we find
\bqn
\lb{6.7a}
{E_{1}}' &=& \frac{1}{\left(V_{ss} - x\right)^{2} - 1}
 \left(A V_{1}
+ B E_{1}\right),\\
\lb{6.7b}
{V_{1}}' &=& \frac{1}{\left(V_{ss} - x\right)^{2} - 1}
\left(C  V_{1}
+ D E_{1}\right),
\eqn
where
\bqn
\lb{6.8}
A &\equiv& \left({V_{ss}}' - k\right) - \left(V_{ss} - x\right){E_{ss}}' -
\frac{1}{(1-k)x^{2}}e^{E_{ss}},\nb\\
B &\equiv& \left(V_{ss} - x\right)\left[k  - 
\frac{1}{(1-k)x^{2}}  e^{E_{ss}}\right],\nb\\
C &\equiv& {E_{ss}}' -  \left(V_{ss} - x\right)
\left[\left({V_{ss}}' - k\right)
+  \frac{1}{(1-k)x^{2}}e^{E_{ss}}\right],\nb\\
D &\equiv& \frac{ \left({V_{ss}} - x\right)^{2}}
{(1-k)x^{2}}e^{E_{ss}} - k.
\eqn
In the following we shall consider the asymptotic behavior of $E_{1}$ and
$V_{1}$ at $x = x_{c}$ and $x = \infty$ separately. First, 
in the neighborhood of $x_{c}$ we expand them as
\bqn
\lb{6.9}
E_{1} &=& E_{1}^{0} + E^{1}_{1}\left(x - x_{c}\right) 
 + R_{E}^{1}(x, x_{c}),\nb\\
V_{1} &=& V_{1}^{0} + V^{1}_{1}\left(x - x_{c}\right)   
+ R_{V}^{1}(x, x_{c}).
\eqn
Substituting the above expressions into Eqs.(\ref{6.7a}) and (\ref{6.7b})
we find that
\bqn
\lb{6.10a}
E^{0}_{1} &=& \frac{(1-k)\left[(k-2)x_{c} - 2\right] -2}
{k(1-k)x_{c} + 2}V^{0}_{1}, \nb\\
E^{1}_{1} &=& - V^{1}_{1} + k E^{0}_{1} + \frac{1+x_{c}}{x_{c}}V^{0}_{1},
\;\; ({\mbox{Type 1}}),
\eqn
for type 1 solutions, and 
\bqn
\lb{6.10b}
E^{0}_{1} &=& \frac{k\left[(1-k)x_{c} - 2\right]}
{k(1-k)x_{c} + 2}V^{0}_{1}, \nb\\
E^{1}_{1}  &=&  - V^{1}_{1} + k E^{0}_{1} - \frac{1 }{x_{c}}V^{0}_{1},
\;\; ({\mbox{Type 2}}),
\eqn
for type 2 solutions, where $V^{0}_{1}$ and $V^{1}_{1}$ are arbitrary 
constants in both  types.
Applying the boundary conditions (\ref{6.3a}) and (\ref{6.3ab}) to
the above solutions we find that  
\bq
\lb{6.11}
E^{0}_{1} = V^{0}_{1} = 0, \;\;\; 
E^{1}_{1} = - V^{1}_{1} = - W_{0},
\eq
for a  given $k$. 

On the other hand, setting $\epsilon
\equiv 1/x$ we can expand $F_{1}$ at the neighborhood of $\epsilon = 0$
as,
\bqn
\lb{6.12}
E^{1} &=& e^{0}_{1} + e^{1}_{1} \epsilon + O\left(\epsilon^{2}\right),\nb\\
M^{1} &=& m^{0}_{1} + m^{1}_{1} \epsilon + O\left(\epsilon^{2}\right),\nb\\
V^{1} &=& v^{0}_{1} + v^{1}_{1} \epsilon + O\left(\epsilon^{2}\right).
\eqn
Inserting these expressions and the ones given by (\ref{6.5}) into
 Eqs.(\ref{6.2a})-(\ref{6.2d})  we find that all the  
coefficients, $f^{i}_{1}$, are zero, unless $k = 0$, for which we have
\bqn
\lb{6.13a}
m^{0}_{1} &=&  D_{\infty}\left(e^{0}_{1} - v^{0}_{1}\right),  \nb\\
e^{1}_{1} &=& 0, \;\;\; v^{1}_{1} = - D_{\infty} e^{0}_{1}, \nb\\
m^{1}_{1} &=&  D_{\infty}
            \left(D_{\infty} - V_{\infty} e^{0}_{1}\right)e^{0}_{1},
	    \;\;  ({\mbox{Type 1}}), 
\eqn
for type 1, and
\bqn
\lb{6.13b}
m^{0}_{1} &=&  D_{\infty}\left(e^{0}_{1} - v^{0}_{1}\right),  \nb\\
e^{1}_{1} &=& 0, \;\;\; v^{1}_{1} = - D_{\infty} e^{0}_{1}, \nb\\
m^{1}_{1} &=&  D_{\infty}
            \left(D_{\infty} - V_{\infty} e^{0}_{1}\right)e^{0}_{1},
	    \;\;  ({\mbox{Type 2}}), 
\eqn
for type 2, where $e^{0}_{1}$ and $v^{1}_{1}$ are arbitrary. However,
in each of the cases  conditions (\ref{6.3b}) are satisfied  
for any given  $k$.

In review all the above, we conclude that {\em the considerations of the 
perturbations outside the sonic line $x = x_{c}$ and their matching to 
the ones obtained along the line do not impose any additional conditions 
on the spectrum, $k$, of the kink perturbations, obtained by considering
the linear perturbations only along the sonic line}.

\section{Discussions and Concluding Remarks}  

In this paper we first gave a general review over the weak discontinuities
of a fluid along a sonic line by providing a concrete example, and showed
how such a discontinuity can propagate along the line. Then, in Sec. II
we summarized the  main properties of the self-similar solutions of an 
isothermal spherical self-similar flow. In particular, we investigated the 
discontinuities of these solutions across the sonic line, which is useful and
necessary for the studies of the kink stability to be considered in the 
following sections. In Sec. III, using
distribution theory we first developed a general formula of perturbations
in the background of a self-similar solution. For such a development, we
assumed that the background is at least $C^{2}$ outside of the
sonic line and $C^{0}$ across it. The field equations were
divided into three different sets, one in the region $x > x_{c}$, 
one in the region $x < x_{c}$, and the other along the sonic line 
$x = x_{c}$. Any perturbations must satisfy all of these equations.
In Sec. IV, following Ori and 
Piran \cite{OP88} we considered the differential equations along the sonic 
line and found their explicit solutions. When the background is smooth across
the sonic line, we re-derived the results of Ori and Piran. When the background is 
not smooth, we found the stability criterion for the self-similar solutions.
In particular, if the external solution is type 1 and the internal
is type 2, the solution is  stable for $- 2 < x_{c} < -1$, and not stable
for $x_{c} \le -2$. If the external solution is type 2 and the internal
is type 1, it is just the other way around. That is, it is  stable for 
$x_{c} < -2$, and  not stable for $- 2 \le x_{c} < -1$. 

Up to this point, it was not clear  whether the solutions of the 
perturbations obtained along the sonic line can be matched to the ones 
outside the sonic line. Since the perturbations in the region
$|x| < |x_{c}|$ is assumed to be  zero identically, the ones in the 
region $|x| > |x_{c}|$ are necessarily non-zero. Otherwise, 
it is impossible to get a non-zero discontinuity across the sonic line. 
To answer the above question, we studied the linear perturbations 
in the region $|x| > |x_{c}|$, by taking the solutions 
obtained along the sonic line as the boundary conditions. By properly imposing 
other boundary conditions at the spatial infinity, we were able to show 
that there always exist linear perturbations in the region $|x| > |x_{c}|$,
which satisfy all the boundary conditions for a given $k$, where $k$ is 
the spectrum of the perturbations obtained along the sonic line. As a result, 
the studies of perturbations along the sonic line carried out before are 
consistent with the complete treatment of the problem in the whole spacetime.
In particular, the spectrum of perturbations obtained along the sonic line
remains the same even after the perturbations in the whole spacetimes
are considered.

Finally, we note that it would be very interesting to generalize the above 
studies to the relativistic case, a subject that is under our current
investigations.

\section*{Acknowledgments}

The authors would like to express their gratitude to R.-G. Cai, M.
W. Choptuik,  and T. Harada  for valuable discussions and suggestions.


\end{document}